\documentclass[twocolumn,graphics,floatfix,a4paper,prl]{revtex4-1}
\usepackage{color}
\usepackage[utf8]{inputenc}
\usepackage{graphicx}
\usepackage{amsmath}
\usepackage{amssymb}
\usepackage{longtable}
\usepackage{pdfpages}

\usepackage{hyperref}
\hypersetup{
  pdfnewwindow=true, colorlinks=true,
  linkcolor=blue, anchorcolor=blue,
  citecolor=blue, filecolor=blue,
  menucolor=blue, urlcolor=blue}

\newcommand{\comment}[1]{\textit{}}
\newcommand{\bit}{\begin{itemize} \setlength{\itemsep}{0ex} \setlength{\topsep}{0ex} } 
\newcommand{\eit}{\end{itemize}}
\newcommand{\be}{\begin{equation}}
\newcommand{\ee}{\end{equation}}
\newcommand{\bea}{\begin{eqnarray}}
\newcommand{\eea}{\end{eqnarray}}
\newcommand{\ba}{\begin{align}}
\newcommand{\ea}{\end{align}}
\newcommand{\SKIP}[1]{}

\providecommand{\bk}{\ensuremath{{\bf k}}}
\providecommand{\kp}{\ensuremath{{\bf k}\cdot{\bf p}}}

\providecommand{\inassb}{I\lowercase{n}A\lowercase{s}$_{1-x}$S\lowercase{b}$_x$}
\providecommand{\inassbh}{I\lowercase{n}A\lowercase{s}$_{0.5}$S\lowercase{b}$_{0.5}$}

\begin{document}

\title{Topological Phases in \inassb{}:\\ From Novel Topological Semimetal to Majorana Wire}
\author{Georg W. Winkler$^1$}
\author{QuanSheng Wu$^1$}
\author{Matthias Troyer$^1$}
\author{Peter Krogstrup$^2$}
\author{Alexey A. Soluyanov$^{1,3}$}
\affiliation{$^1$Theoretical Physics and Station Q Zurich, ETH Zurich, 8093 Zurich, Switzerland}
\affiliation{$^2$Center for Quantum Devices and Station Q Copenhagen, Niels Bohr Institute, University of Copenhagen, 2100 Copenhagen, Denmark}
\affiliation{$^3$Department of Physics, St. Petersburg State University, St. Petersburg, 199034 Russia}
\date{\today}

\begin{abstract}
Superconductor proximitized one-dimensional semiconductor nanowires with strong spin-orbit interaction (SOI)  are at this time the most promising candidates for the realization of topological quantum information processing. In current experiments the SOI originates predominantly from extrinsic fields, induced by finite size effects and applied gate voltages. The dependence of the topological transition in these devices on microscopic details makes scaling to a large number of devices difficult unless a material with dominant intrinsic bulk SOI is used. Here we show that wires made of certain ordered alloys \inassb{} have spin splittings up to 20 times larger than those reached in pristine InSb wires. In particular, we show this for a stable ordered CuPt-structure at $x = 0.5$, which has an inverted band ordering and realizes a novel type of a topological semimetal with triple degeneracy points in the bulk spectrum that produce topological surface Fermi arcs. Experimentally achievable strains can drive this compound either into a topological insulator phase, or restore the normal band ordering making the CuPt-ordered \inassbh{} a semiconductor with a large intrinsic linear in $k$ bulk spin splitting. 
\end{abstract}

\maketitle
In recent years, a range of topological phases have been realized in materials, ranging from topological insulators~\cite{hasan_colloquium, qi-zhang} (TIs) and semimetals~\cite{turner_semimetals_2013,Vishwanath,wang_cd3as2,cd3as2_experiment} (TSMs) to superconductors~\cite{volovik_book,flensberg_superconductivity} (TSCs). The non-trivial topology of the ground state wavefunctions in these phases causes a variety of phenomena in such materials ranging from topologically protected metallic surface or edge states in TIs~\cite{hasan_colloquium, qi-zhang} and Fermi arcs and anomalous magnetotransport in TSMs~\cite{NIELSEN198120,Vishwanath,qi_transport,son_spivak}, to quasiparticles with non-Abelian particle statistics~\cite{ivanov,reed_green,Yazdani,Lutchyn,Oreg,alicea_prb,Alicea,mourik_majorana_2012} in TSCs, which could be used for topological quantum computation~\cite{top_comp1,top_comp2}.  

Arguably the  simplest scheme for realizing non-Abelian statistics in a solid-state device is based on manipulating Majorana zero modes (MZMs) in networks of semiconductor wires. MZMs were predicted to appear at the ends of spin-orbit coupled wires subject to a parallel magnetic field, proximity coupled to an $s$-wave superconductor. Experimental observations, consistent with the theory, were reported for InAs and InSb zincblende nanowires~\cite{das_majorana_2012,xu_majorana_2012,mourik_majorana_2012,CM_Majorana}. 

The stability of MZMs in such a setup depends greatly on the size of the spin-orbit splitting (SOS) of the conduction band.
 SOS is very small in bulk zincblende semiconductors~\cite{dresselhaus_spin_zincblende} and the realization of the MZMs thus relies on the externally induced Rashba SOS~\cite{Rashba}, which is estimated to be of the order of 1~meV~\cite{splitting_insb_delft,alexey_wire}.
This value is very small compared to the bulk splitting in some recently discovered compounds~\citep{BiTeI,GeTe,BiTeCl,Chulkov}. However, most of these materials are not suitable for realizing MZMs within the above scenario, while for others such experiments appear to be challenging. It is thus desirable to understand if large values of bulk SOS can be achieved within the III-V materials class, used in most  experiments at this time. A bulk SOS dominating the Rashba contributions would also make realization of MZMs far less sensitive to particular microscopic details of a specific device.  

\begin{figure}[t]
\includegraphics[width = 0.5 \linewidth]{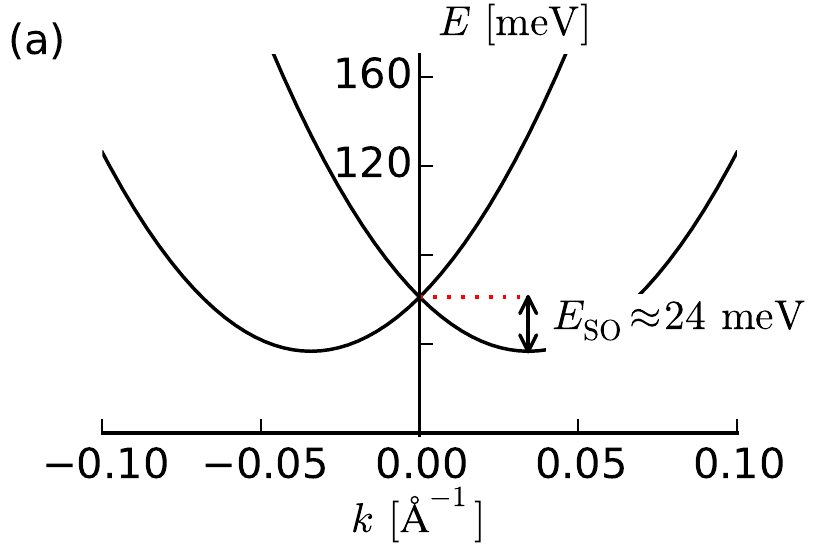}\includegraphics[width = 0.5 \linewidth]{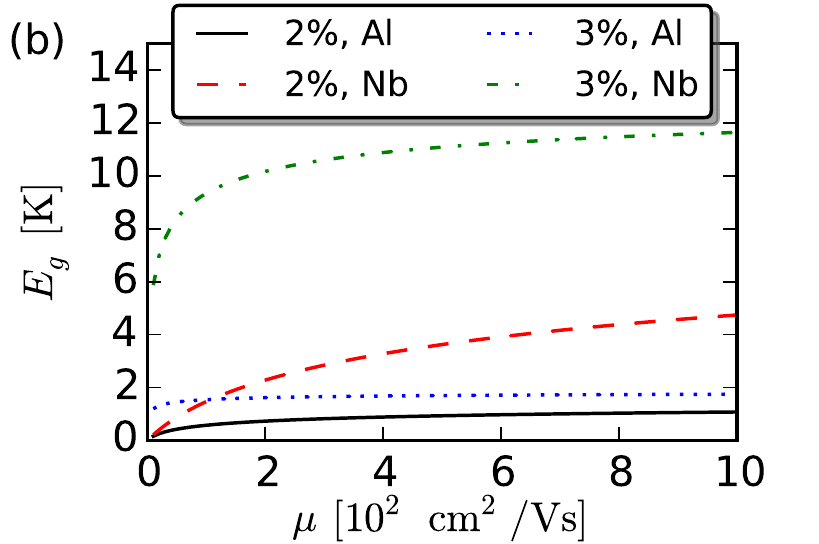}
\caption{(a) Conduction bands of $-3\,\%$ strained \inassbh{} plotted in the $(1\bar 10)$ direction. (b) Disorder renormalized quasiparticle gap $E_g$ in the TSC state as a function of the mobility plotted for 2\% and 3\% strain for a $(1\bar 10)$ wire in proximity to Al or Nb.}
  \label{fig:splitting}  
\end{figure}

In this Letter we argue that certain {\it ordered} alloys with composition InAs$_{1-x}$Sb$_x$ have sizable SOS and provide an optimal material for the realization of MZMs.
For ternary alloys to give an advantage over fixed binary \mbox{III-V} compounds, SOS should be enhanced, ideally while preserving the high carrier mobility. We thus focus our attention on ordered superlattice structures.
We find, in particuar, that a CuPt-ordering with $x=0.5$  is energetically stable and 
hosts a {\em novel TSM phase}, which is identified as an interpolation of the established Dirac~\cite{wang_dirac_2012,nagaosa,wang_cd2as3} and Weyl~\cite{Vishwanath, volovik_book} TSMs.
Moreover, the TSM phase can be tuned either into a TI or normal insulator phase by application of strain. We also find that for the latter the spin-orbit energy $E_{\mathrm{SO}}$~\footnote{The definition of $E_{\mathrm{SO}}$ is given in Fig.~\ref{fig:splitting}(a)} can be as large as 24~meV. Experimental evidence for the CuPt-ordering of InAs$_{0.5}$Sb$_{0.5}$ exists~\cite{stringfellow_1989,stringfellow_1991,kurtz_1992,nanowire_cupt,suchalkin_ordering_2015}, and we argue that nanowires of this structure can be grown with molecular beam epitaxy. 

The SOS of the conduction band in zincblende structures is at most cubic in $k$ around the $\Gamma$-point~\cite{dresselhaus_spin_zincblende}. In both III-V materials, the conduction band has a minimum at $\Gamma$  and an $s$-like character dictated by $T_d$ symmetry. A SOS linear in $k$ can thus only be achieved by breaking the $T_d$ symmetry. We ask the question if there is a modification to the III-V materials, such that a bulk linear in $k$ SOS is achieved.

\begin{figure}
\includegraphics[width =  \linewidth]{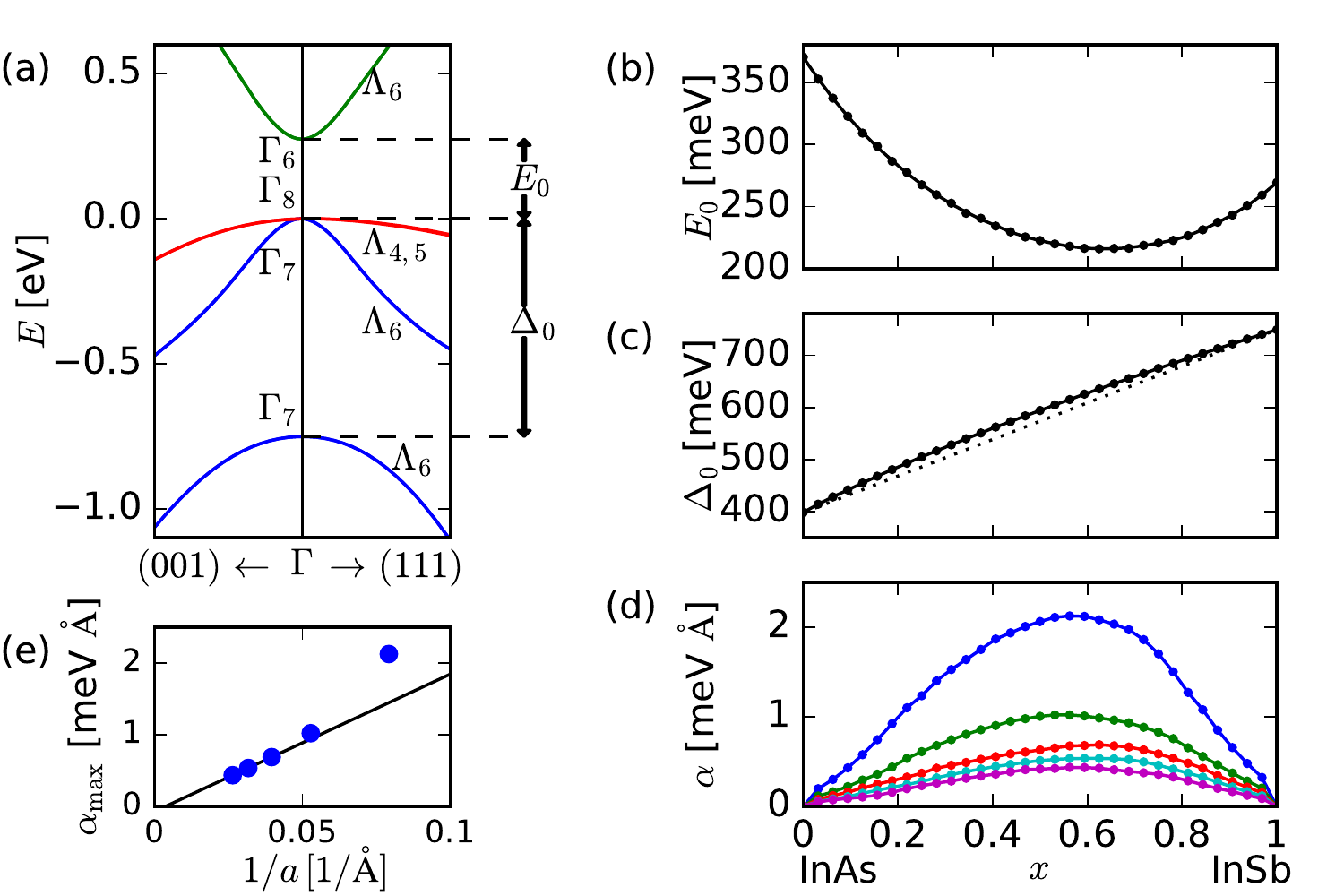}
\caption{ (a): First-principles band structure of zincblende InSb around the $\Gamma$ point, plotted up to $|{\bf k}|=0.1\,${\AA}$^{-1}$. Apart from the $\Lambda_{4,5}$ representations each plotted band is two-fold degenerate. The next panels  illustrate the change in the band structure of InAs$_{1-x}$Sb$_x$ as a function of $x$ obtained from ETB supercell calculations:  (b) the fundamental gap $E_0(x)$, (c) the spin-orbit gap $\Delta_0(x)$, with the linear dotted line as a guide to show the bowing, and (d) the Rashba coefficient $\alpha(x)$ of Eq.~\ref{eq:alpha} for the SOS in the (110)-direction. Various supercell sizes are shown:  64~(blue), 216~(green), 512~(red), 1000~(cyan) and 1728~(magenta) atoms.  (e) the maximum splitting $\alpha_\mathrm{max}$ with linear extrapolation to an infinitely large cubic supercell.}
  \label{fig:gaps}  
\end{figure}

\paragraph{Disorder in nanoscale structures} --- In the randomly disordered alloy \inassb{} any $T_d$ symmetry breaking terms need to vanish because of averaging. However, in a nanoscale device, like a quantum dot or wire, the averaging will not be complete and a ``bulk'' contribution to the linear in $k$ SOS is expected. We study this effect by simulating randomly disordered cubic supercells of \inassb{}~\footnote{The virtual crystal approximation~\cite{VCA} (VCA), which is a standard approximation for simulating alloying, is not able to reproduce experimentally known features of \inassb{}, such as the nonlinear bowing of the fundamental energy gap $E_0$ and the spin-orbit gap $\Delta_0$ (see Fig.~\ref{fig:gaps}(a))~\cite{chadi_1977,bowing_2008,bowing_2012}.}. The supercell calculations are faciliated within Slater-Koster tight-binding models~\cite{slater_koster} of the sp$^3$s$^*$ type~\cite{vogl_sp3s,klimeck_sp3s,carlo_sp3s}, with parameters derived from first principles bulk calculations employing the HSE06 hybrid functional~\cite{heyd_hybrid_2003,heyd_hybrid_2004,heyd_hybrid_2006,kresse_hse06}. See Supplementary Material~\footnote{See Supplementary Information for a detailed description of (additional) first principles calculations, tight-binding models, derivation of $\kp$ models, details on the topological classification and Land\'e $g$-factor calculation.} for technical details on the calculation.

Figure~\ref{fig:gaps}  (b-c)  show the composition dependence of the band gap $E_0$ and the spin-orbit gap $\Delta_0$. The curves $E_0(x)$ and $\Delta_0(x)$ exhibit the correct bowing (non-linearity), and are in good agreement with recent experiments~\cite{bowing_2008,bowing_2012,suchalkin_gap_2016}. The linear in $k$ SOS of the lowest conduction band is quantified by fitting the band structure calculations to an effective Hamiltonian 
\begin{equation}
  E^{\pm}_c(k) = \frac{\hbar^2 k^2}{2 m^*} \pm \alpha \,k .
  \label{eq:alpha}
\end{equation}
The parameter $\alpha$ plays the role of the Rashba parameter in standard models for Majorana wires~\cite{Alicea, Lutchyn, Oreg}. The magnitude of this coefficient varies with the $k$-space direction, and its dependence on disorder realizations is shown in Fig.~\ref{fig:gaps}(d) for the (110) direction~\footnote{Which is the direction of the maximal SOS in pure zincblende semiconductors for small $k$~\cite{Dresselhaus,zunger_splitting,alexey_wire}.}. As expected, $\alpha$ decreases due to averaging when the supercell size is enlarged, which is also shown in Figure~\ref{fig:gaps}(e). As a consequence, wires or quantum dots of randomly alloyed InAs$_{1-x}$Sb$_{x}$ have a universal bulk contribution to the SOS that depends strongly on the size of the device.

\paragraph{CuPt-ordered structure} --- In contrast to the disordered configurations above, $T_d$ symmetry breaking by alloy ordering is nonvanishing for arbitrary system size. Since Fig.~\ref{fig:gaps}(d) shows that $\alpha$ is maximized in the vicinity of $x \approx 0.6$ we focus on small ordered supercells with $x=0.5$ as good candidates for realization of large SOS, maximizing $E_{\mathrm{SO}}=\frac{m^* \alpha^2}{2 \hbar^2}$.
In particular, we considered three types of ordering reported in experiments on III-V ternary alloys~\cite{stringfellow_1991}: ordering in \{100\} planes (CuAu-I or L1$_0$ structure), ordering in \{210\} planes (chalcopyrite or E1$_1$ structure) and ordering in \{111\} planes (CuPt or L1$_1$ structure).
The CuPt-type ordering is energetically stable in \inassbh{}~\cite{Note3} and has been experimentally observed under various growth conditions~\cite{stringfellow_1989,stringfellow_1991,kurtz_1992,nanowire_cupt,suchalkin_ordering_2015}. We find that this structure also has the biggest enhancement of the SOS (see Supplementary Material~\cite{Note3}). We thus consider this type of ordering in the following. 

\begin{figure}
  \center
\hspace{0.05 \linewidth}\includegraphics[width =  0.34 \linewidth]{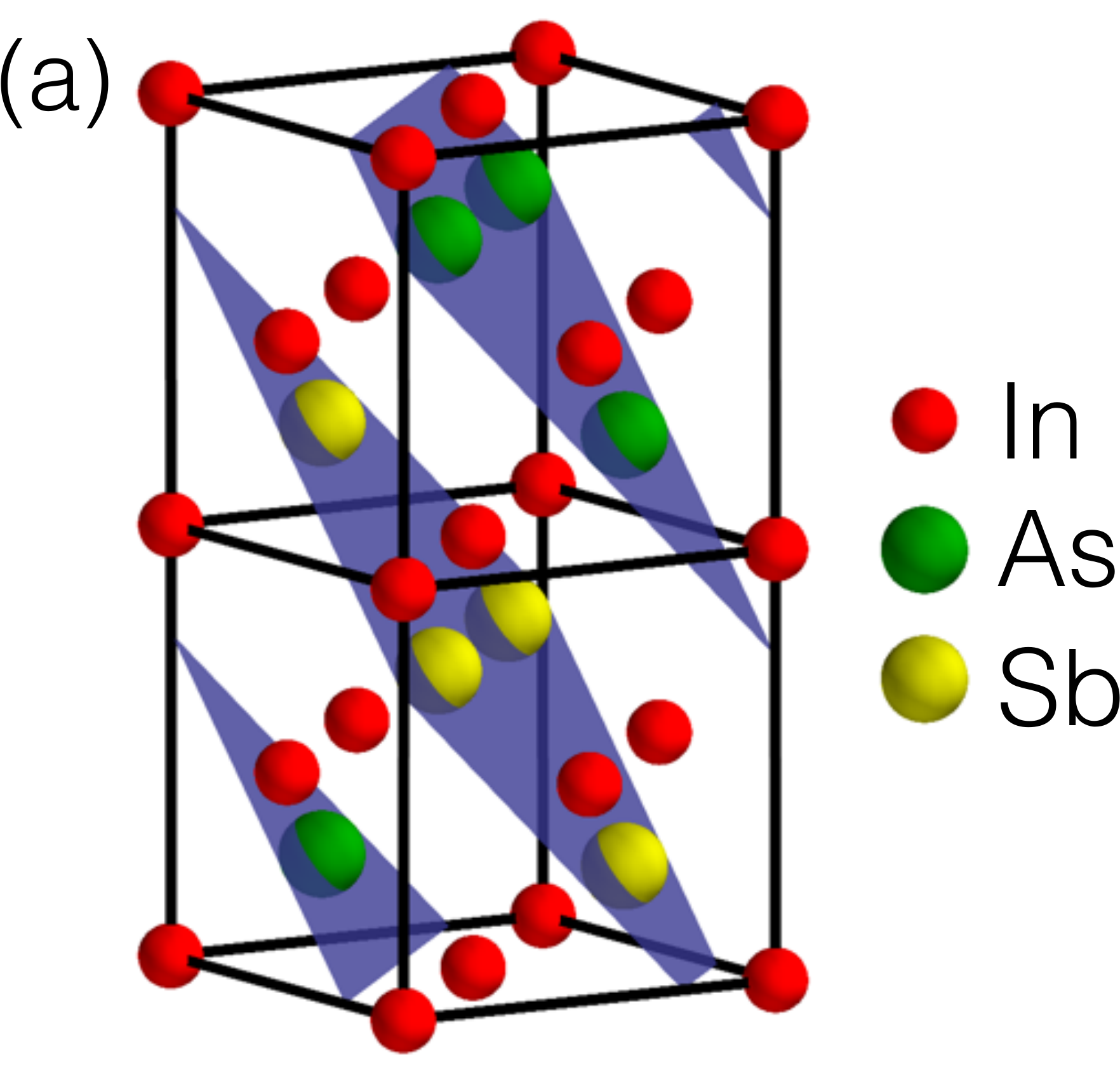}\hspace{0.05 \linewidth}\includegraphics[width =  0.5 \linewidth]{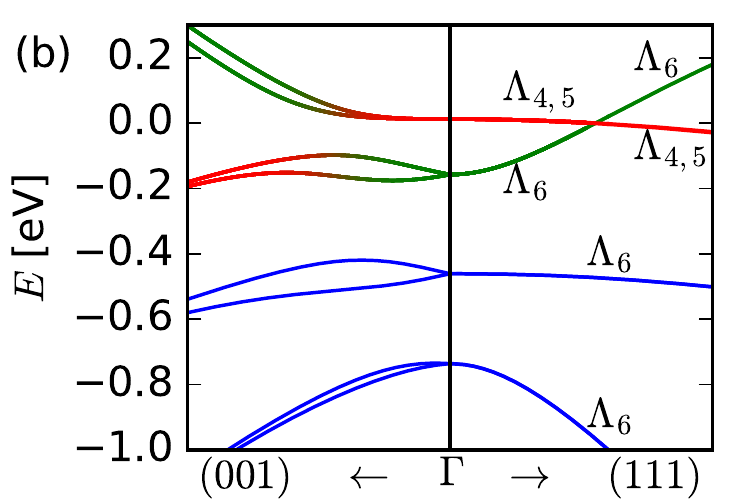}\hspace{0.05 \linewidth}
\caption{(a) Crystal structure of CuPt-ordered \inassbh{}. (b) Band structure of CuPt-ordered \inassbh{} around $\Gamma$ (plotted up to $|{\bf k}|=0.1\,\mathrm{\AA}^{-1}$). The colors correspond to the original band ordering in Fig.~\ref{fig:gaps}.}
  \label{fig:cupt}  
\end{figure}

In the CuPt-ordered \inassbh{} structure, shown in Fig.~\ref{fig:cupt}(a), atoms of As and Sb are arranged in  \{111\} planes that grow in the alternating order In-As-In-Sb. This ordering reduces the $T_d$ (space group \#216) symmetry of the pristine zincblende compounds to $C_{3v}$ (space group \#160), that has a three-fold rotational axis coinciding with the (111) direction, and three vertical mirror planes that contain the symmetry axis and are rotated by $2\pi/3$ relative to each other. Both the ionic positions and lattice vectors were relaxed using the HSE06 hybrid functional to get the lowest energy structure. We found the deviation from the cubic structure to be smaller than 0.1\%~\cite{Note3}.

The band structure of CuPt-ordered InAs$_{0.5}$Sb$_{0.5}$ obtained with HSE06 is shown in Fig.~\ref{fig:cupt}(b).  The little group of $k$-points on the (111)-axis in the CuPt-structure is $C_{3v}$ as in the zincblende structure, hence the same symmetry label $\Lambda$ is used for the bands in Figs.~\ref{fig:gaps}(a) and~\ref{fig:cupt}(b). 
While the $\Lambda_{4,5}$ bands have a very small, linear in $k$, splitting, the bands of the $\Lambda_6$ representation are doubly degenerate. Note that the ordering of the valence and conduction bands at $\Gamma$ is interchanged for the two structures, resulting in a {\em band inversion} in the CuPt-structure. We find the band inversion to be stable against deviations from the CuPt-order, in particular we find that CuPt-ordered InAs$_{0.67}$Sb$_{0.33}$ and InAs$_{0.33}$Sb$_{0.67}$ still show a strong band inversion~\cite{Note3}.

\begin{figure}
\includegraphics[width = \linewidth]{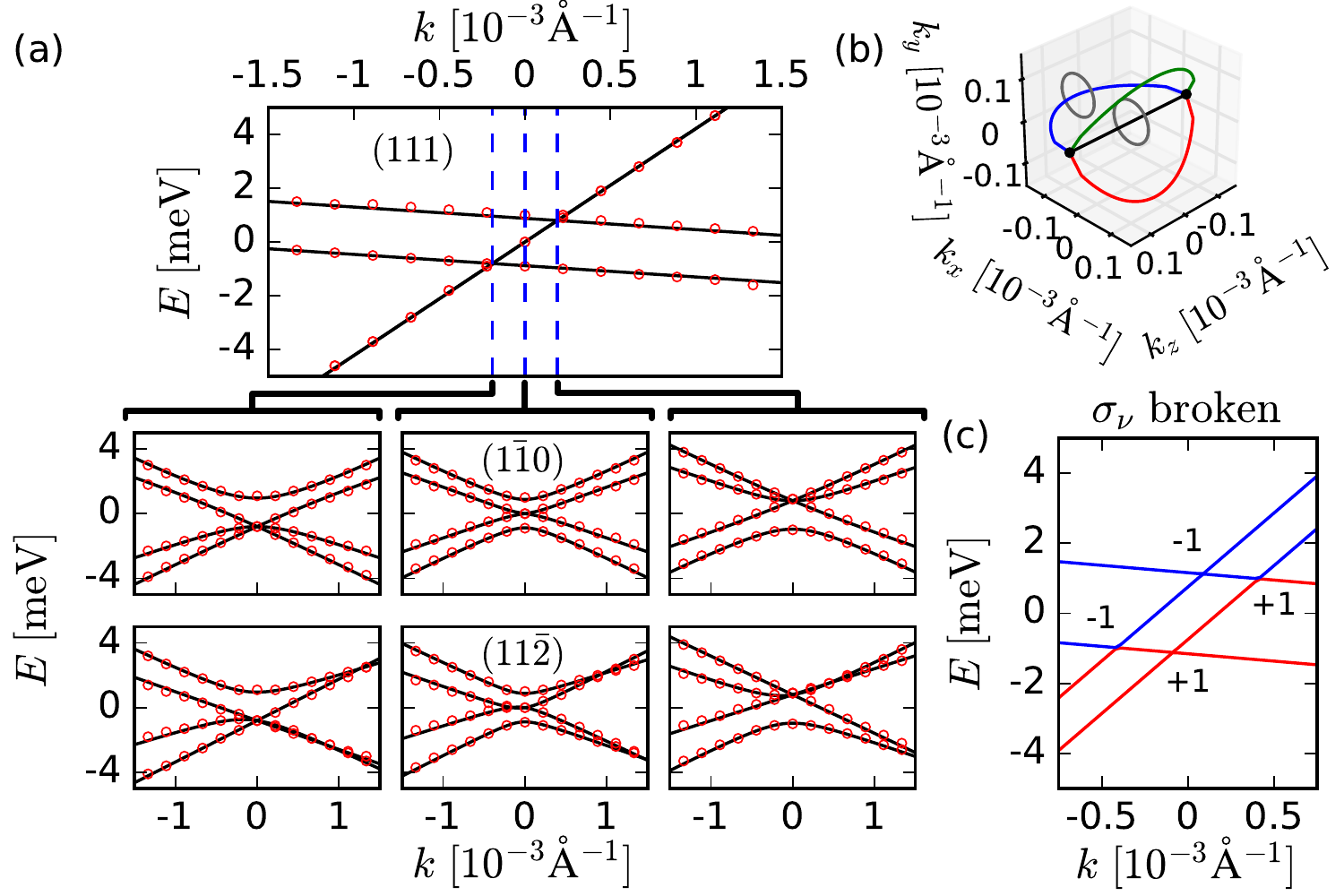}
\caption{(a) Zoom in of the TP for the three orthogonal $k$-directions and at three different values of of momentum along the (111) direction. Band structures obtained from both first-principles calculations (red dots) and the $\kp$ model of Eq.~\eqref{eq:kp} (solid line) are shown. (b) Nodal lines (red, green and blue) connecting the two TPs (black dots). The grey circles indicate paths for Berry phase calculation.
  (c) The $\kp$ band structure for the  (111)-direction with broken mirror symmetries. Four Weyl points and their chiral charges are shown. }
  \label{fig:crossing}  
\end{figure}

\paragraph{A novel TSM} --- In the inverted band structure the $\Lambda_6$ and $\Lambda_{4,5}$ bands cross (see Fig.~\ref{fig:crossing}(a)) to form a new type of a TSM. While theoretical evidence for the band inversion in ordered InAs$_{0.5}$Sb$_{0.5}$ was reported previously~\cite{zunger_inversion_1991}, the topology of this semimetal phase was overlooked. 

Here the crossings are protected by the $C_3$-rotational symmetry. For inversion-symmetric materials $C_3$ is known to stabilize Dirac points on the high-symmetry axis~\cite{wang_dirac_2012, turner_semimetals_2013, nagaosa}.
The inversion symmetry is absent in InAs$_{0.5}$Sb$_{0.5}$, so that two of the four bands that would form a Dirac point are gapped.
Along the (111)-axis, the two singly degenerate bands $\Lambda_{4,5}$ each cross with the doubly degenerate $\Lambda_{6}$ forming triply degenerate crossings, or triple points (TPs). In the other two directions each TP splits into two linearly dispersing bands and a quadratically dispersing one (see Fig.~\ref{fig:crossing}(a)). Previously, TPs have been discussed in the context of Bernal-stacked graphite with neglected SOI~\cite{mikitik_sharlai,nexus1,nexus2}, spin-1 quasiparticles in two dimensions~\cite{2D_TP_1,2D_TP_2} and strained HgTe~\cite{zaheer_hgte}. The TP we find is furthermore accompanied by four Weyl nodal lines~\cite{nodal_line} in the vertical mirror planes, degenerate lines between the second and third band in Fig.~\ref{fig:crossing}(a), each protected by a Berry phase of $\pi$ (see Fig.~\ref{fig:crossing}(b)).


%
\begin{table}[b]
  \begin{tabular}{ c  c  c  }
    \hline
    \hline
    $E_0$(meV) & $A$(eV \AA) &  $B$(eV \AA) \\
    \hline
    0.88 & -0.42 & 4.22 \\
    $C$(eV \AA) & $D$(eV \AA) & $F$(eV \AA) \\
    \hline
    0.78 & $1.25-i1.51$ & 2.14
\end{tabular}
\caption{Parameters of the fitted $\kp$ model.}
\label{tab:par}
\end{table}
%

In the vicinity of the crossing point the band structure can be described by the following $\kp$ model (see Supplementary Material~\cite{Note3})
\begin{equation}
  \scriptsize
  H_{\kp} = \begin{pmatrix}
    E_0 + A k_z & 0 & D^{\phantom{*}} k_y &  \phantom{-}D^{\phantom{*}}k_x \\
    0 & -E_0+A k_z & F^{*}k_x & -F^* k_y \\
    D^{*} k_y &  \phantom{-}F^{\phantom{*}} k_x & B k_z + C k_x & \phantom{-}C^{\phantom{*}}k_y \\
    D^* k_x & -F^{\phantom{*}}k_y & C^{\phantom{*}}k_y & B k_z-C k_x
  \end{pmatrix}.
  \label{eq:kp}
\end{equation}
For InAs$_{0.5}$Sb$_{0.5}$ the values of the parameters obtained from fits  to the first-principles calculations are listed in Tab.~\ref{tab:par}, and the fit is shown in Fig.~\ref{fig:crossing}(a). The momentum $(k_x,k_y,k_z)$ here is given relative to the crossing point ${\bf k}_c = (0,0,k_c=0.0646\,\mathrm{\AA}^{-1})$, and $k_z$ is taken to be in the (111)-direction and $k_y$ in the $(1\overline{1}0)$-direction.

While a detailed description and topological classification of this novel TSM phase will be provided elsewhere~\cite{crossing_point}, we outline the proof of the topological origin of this phase here.
An illustrative verification of the topological origin of this phase is obtained by breaking the mirror symmetry $\sigma_v$ of the $C_{3v}$ group. The doubly degenerate band $\Lambda_{6}$ splits into two bands, and four crossings are formed as shown in Fig.~\ref{fig:crossing}(c). All the four crossings represent Weyl points, with their chiral charges shown in the figure~\cite{Note3}. On the other hand, if all inversion symmetry breaking terms are tuned to zero one obtains a Dirac TSM and thus the TP TSM can be seen as an interpolation of Dirac and Weyl TSMs.

\begin{figure}
\includegraphics[width = \linewidth]{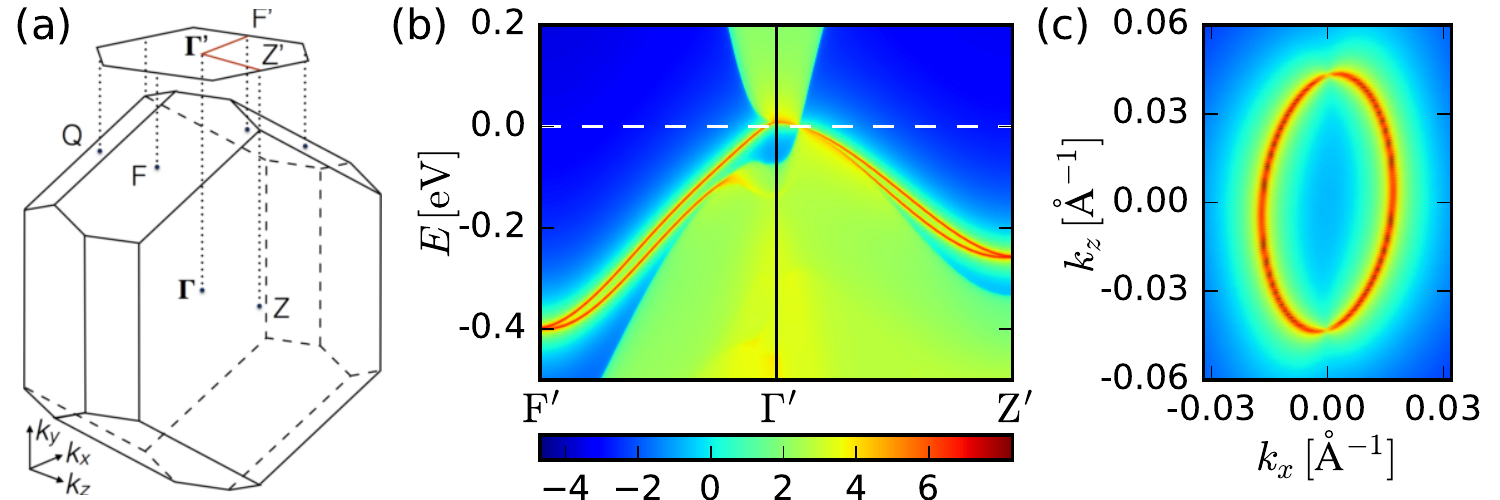}
\caption{(a) Brillouin zone and surface projection of CuPt-ordered \inassbh{}. (b) Surface density of states for the $(1\bar 1 0)$ surface. (c) Topological Fermi arcs on the $(1\bar 1 0)$ surface. }
  \label{fig:surfdos}  
\end{figure}

The presence of Dirac or Weyl points in the bulk spectrum of metals is associated with the appearance of Fermi arcs in the surface spectrum~\cite{weng_dai_bernevig, alexey_weyl, wang_TaAs, hasan_nat_comm,wang_dirac_2012, wang_cd2as3,Xu_dirac_arpes}. Fig.~\ref{fig:surfdos} shows the surface density of states, obtained by the iterative Green's function method~\cite{iterative_greens_function}. Topological surface states forming the two Fermi arcs that connect the two pairs of TPs at opposite $\bk$ are clearly visible, similar to the ones found in Dirac TSMs. Furthermore, we show in the Supplementary Material~\cite{Note3} that the Landau level spectrum is gapless for a magnetic field aligned parallel to the symmetry axis, hinting at anomalous transport in the presence of magnetic fields~\cite{NIELSEN198120,qi_transport,son_spivak}.

\begin{figure}[b]
\includegraphics[width = \linewidth]{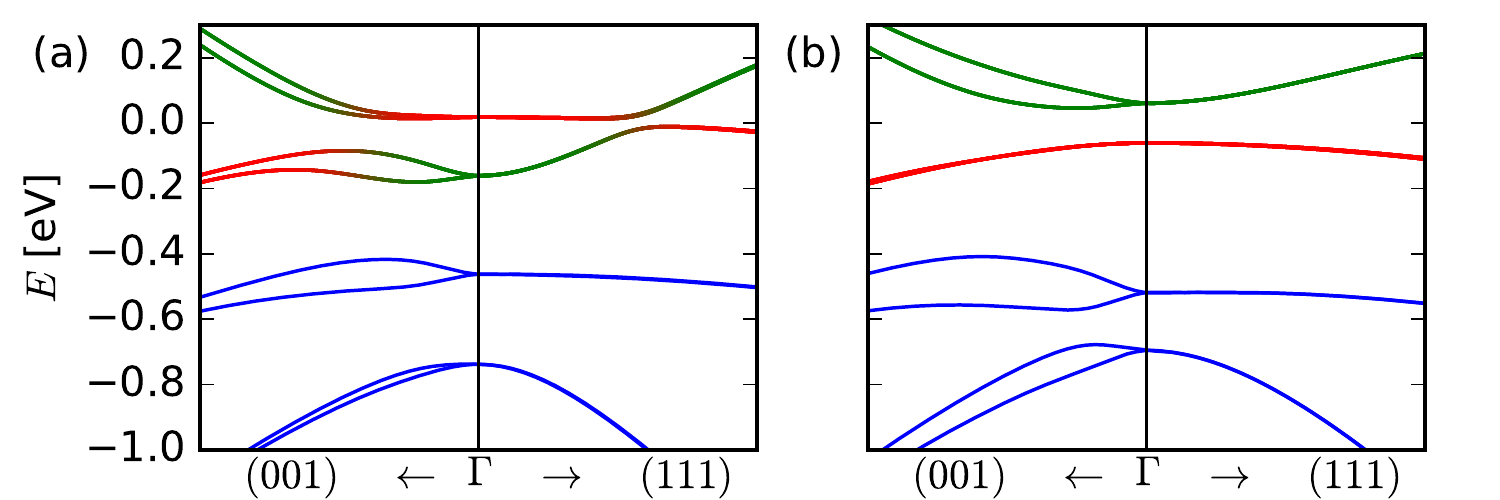}
\caption{Valence and conduction energy bands near the $\Gamma$ point, plotted up to $|{\bf k}|=0.1\,\mathrm{\AA}^{-1}$, for CuPt-ordered \inassbh{} under strain. The colors correspond to the original band ordering of Fig.~\ref{fig:gaps}.
  (a) 1\% compressive strain in the (001)-direction. (b) 3\% compressive strain applied in the (111)-direction.}
  \label{fig:strain}  
\end{figure}

\paragraph{Optimizing the structure for MZMs} --- Having established the existence of a novel TSM phase in InAs$_{0.5}$Sb$_{0.5}$, we now return to our original purpose of finding an optimal structure for MZM realization. Upon breaking the $C_3$ symmetry by strain (e.g. (001)-strain) the degeneracy at the TPs is lifted and the system becomes a semiconductor. 
The inverted band structure makes it a strong 3D TI in this case, as verified by computing the $\mathbb{Z}_2$ topological invariant~\cite{gresch}.
Figure~\ref{fig:strain} (a) shows the HSE06 band structure for 1\% compressive strain in (001) direction, when all the symmetries of the $C_{3v}$ structure are broken. 
In this configuration MZMs can appear in proximitized wires of TIs as discussed by several works~\cite{cook_ti_wire1,cook_ti_wire2}. 

The ordinary band ordering in CuPt-ordered InAs$_{0.5}$Sb$_{0.5}$ can be restored by applying a symmetry-preserving compressive (111)-strain of $>$2\%. Such strain values are easily achievable in nanowires due to lattice mismatch~\cite{strain1,strain2}. Moreover, epitaxial semi-super InAs/Al nanowires
were reported to
bend during the growth process, thus being naturally strained asymmetrically along the (111) growth direction~\cite{peter}.

\begin{table}[b]
  \begin{tabular}{ c | r |  c  c  c  c  c}
    \hline
    \hline
  Material  & & $(111)$ & $(110)$ &  $(001)$ & $(1\bar 10)$  & $(11\bar 2)$ \\
  \hline
  \hline

                   & $\alpha$ [eV\,\AA]    & 0.0 &  0.8 & 1.0 & 1.1 & 1.0 \\
                   & $E_\mathrm{SO}$ [meV] & 0.0 & 2.7 & 3.4 &  3.5 & 3.5 \\
\inassbh{}         & $m^*$ [$m_e$] & 0.24  & 0.16 & 0.14 &  0.12 & 0.13 \\
-2\% (111)-strain  & $l_\mathrm{R}$ [\AA]  & $\infty$ & 60 & 55 &  60 & 60 \\
                   & $|g|$                 & 115 & 95 & 65 & 194 & 177 \\
                   & $E_g$ [$K$]           & 0.0 & 1.0 & 1.1 &  1.1 & 1.1  \\

    \hline

                   & $\alpha$ [eV\,\AA]    & 0.0 &  0.8 & 1.2 & 1.4 & 1.4 \\
                   & $E_\mathrm{SO}$ [meV] & 0.0 & 6 & 14 &  24 & 24 \\
\inassbh{}         & $m^*$ [$m_e$]         & 0.47 & 0.37 & 0.41 &  0.47 & 0.50 \\
-3\% (111)-strain  & $l_\mathrm{R}$ [\AA]  & $\infty$ & 25 & 16 &  11 & 11 \\
                   & $|g|$                 & 83 & 69 & 47 & 164 & 161 \\
                   & $E_g$ [$K$]           & 0.0 & 1.5 & 1.7 &  1.7 & 1.8 \\

\end{tabular}
\caption{Linear coefficient $\alpha$ of Eq.~\eqref{eq:alpha} and $E_\mathrm{SO}$ for different $k$-space directions in the CuPt-ordered compound. The spin-orbit precession length $l_\mathrm{R}=\hbar^2/m^* \alpha$ and the Land\'e $g$-factor, for magnetic field parallel to a gate-defined wire in a thin film of 50 nm thickness~\cite{Note3}, is also shown. $E_g$ is calculated using Al as the bulk superconductor and  assuming a mobility of $\mu = 10^3\,\mathrm{cm}^2/\mathrm{Vs}$.}
\label{tab:alpha}
\end{table}

The HSE06 band structure of (111)-strained InAs$_{0.5}$Sb$_{0.5}$ is shown in Fig.~\ref{fig:strain}(b) for a 3\% strain. In this case the conduction bands acquire a sizable linear in $k$ SOS in any direction but (111) (see  Tab.~\ref{tab:alpha}). We find that the Rashba coefficient $\alpha$ is significantly larger than that reported for pristine InSb nanowires~\cite{splitting_insb_delft}, and the corresponding $E_\mathrm{SO}$ can reach values up to 24~meV, as illustrated in Fig.~\ref{fig:splitting}(a) and Tab.~\ref{tab:alpha}. Note, that the SOS considered here is bulk only and will additionally contribute to the Rashba splitting appearing in a confined geometry.

From Tab.~\ref{tab:alpha} one can see that $E_{\mathrm{SO}}$ is large for all the directions orthogonal to the $C_3$-axis. Thus, an optimal SOS is achieved in wires grown in the plane of CuPt-structure atomic layers. This suggests gate-defined wires~\cite{gate_defined} in (111) thin films of InAs$_{0.5}$Sb$_{0.5}$ to be the most advantageous route to increased stability of MZMs. Additionally, the confinement of the quantum well in the (111) direction has a similar effect as strain and can restore the normal band order for thin quantum wells as is the case in HgTe quantum-wells~\cite{bernevig_hgte}.

To give a rough estimate for the realistic value of a TSC gap induced in InAs$_{0.5}$Sb$_{0.5}$, we used the obtained parameters in the effective model analysis of Ref.~\cite{sau_eg_2012} to calculate the disorder renormalized quasiparticle gap $E_g$ in the TSC state. The gap values $\Delta_\mathrm{s}$ in the adjacent bulk superconductor are taken to be  2~K for Al and 15~K for Nb. In Fig.~\ref{fig:splitting}(b) $E_g$ is plotted as a function of the mobility $\mu$, and Tab.~\ref{tab:alpha} lists values of $E_g$ assuming Al as the bulk superconductor and a mobility of $\mu = 10^3\,\mathrm{cm}^2/\mathrm{Vs}$. Even with this moderate electron mobility $E_g$ is almost an order of magnitude larger in InAs$_{0.5}$Sb$_{0.5}$ than what is currently achievable in pure InSb~\cite{splitting_insb_delft}.

Finally, we considered other III-V alloys, for which CuPt-ordering was reported in Ref.~\cite{stringfellow_1991}. We found that InP$_{0.5}$Sb$_{0.5}$ and GaAs$_{0.5}$Sb$_{0.5}$ realize the novel TSM phase reported above for InAs$_{0.5}$Sb$_{0.5}$. Of the compounds with normal band ordering Al$_{0.5}$Ga$_{0.5}$As and Al$_{0.5}$In$_{0.5}$Sb exhibit the largest values of $E_\mathrm{SO}$, which is small compared to InAs$_{0.5}$Sb$_{0.5}$, being of order 0.1~meV~\cite{Note3}.

{\it Acknowledgments.} We would like to thank C. M. Marcus, M. Wimmer and D. Gresch for useful discussions. G. W. W. wants to thank specifically T. Hyart for a correction and useful discussions. This work was supported by Microsoft Research, the European Research Council through ERC Advanced Grant SIMCOFE, and the Swiss National Science Foundation through the National Competence Centers in Research MARVEL and QSIT.

\bibliography{literature}
\newpage
\includepdf[pages=1]{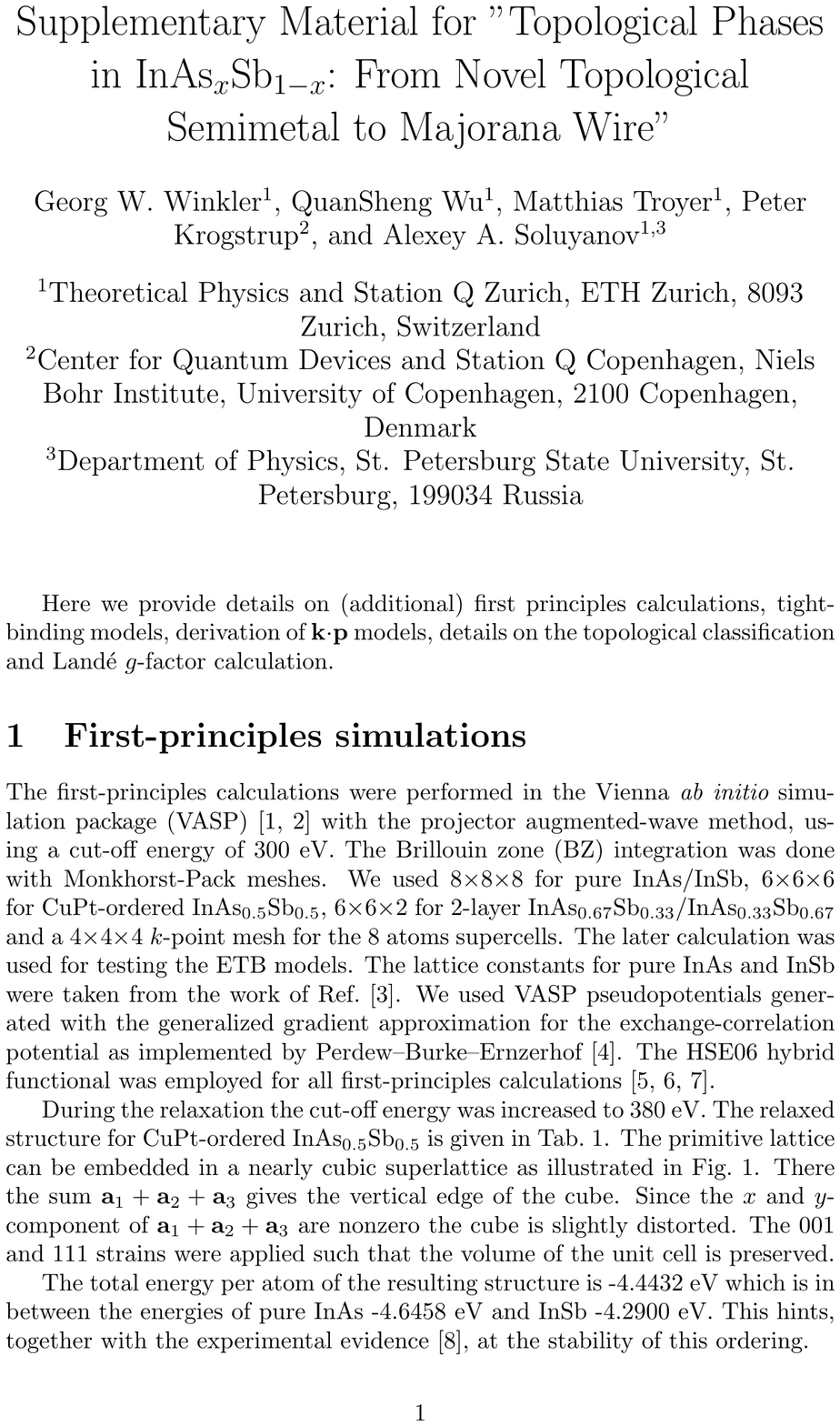}
\mbox{}
\newpage
\includepdf[pages=2]{supplementary_material.pdf}
\mbox{}
\newpage
\includepdf[pages=3]{supplementary_material.pdf}
\mbox{}
\newpage
\includepdf[pages=4]{supplementary_material.pdf}
\mbox{}
\newpage
\includepdf[pages=5]{supplementary_material.pdf}
\mbox{}
\newpage
\includepdf[pages=6]{supplementary_material.pdf}
\mbox{}
\newpage
\includepdf[pages=7]{supplementary_material.pdf}
\mbox{}
\newpage
\includepdf[pages=8]{supplementary_material.pdf}
\mbox{}
\newpage
\includepdf[pages=9]{supplementary_material.pdf}

\end{document}